\documentclass[pra,aps,superscriptaddress,twocolumn,showpacs,nofootinbib]{revtex4-1}

\usepackage{graphicx,color,epstopdf}
\usepackage{amsmath,amsfonts,amssymb,enumerate}
\usepackage{multirow}
\usepackage{hyperref}

\hypersetup{
   colorlinks=true,         
   linkcolor=blue,          
   citecolor=blue,          
   hyperfootnotes=false
}
\usepackage{amsthm}

\def\>{\rangle}
\def\<{\langle}

\def\D{ {\cal D} }

\newcommand{\bra}[1]{\langle {#1} |}
\newcommand{\ket}[1]{| {#1} \rangle}
 
\newcommand{\ketbra}[2]{\ensuremath{\left|#1\right\rangle\!\!\left\langle#2\right|}}

\newcommand{\matrixel}[3]{\ensuremath{\left\langle #1 \vphantom{#2#3} \right| #2 \left| #3 \vphantom{#1#2} \right\rangle}}
\newcommand{\trace}[2]{\mathrm{Tr}_{#1}\left( #2 \right)}
\newcommand{\iden}{\mathbb{I}}
\newcommand{\delt}{\langle \bar{\Delta} \rangle}

\newcommand{\be}{\begin{equation}}
\newcommand{\ee}{\end{equation}}
\newcommand{\norm}[1]{\left\lVert#1\right\rVert}

\newtheorem{thm}{Theorem}
\newtheorem{numbered_theorem}{Theorem}
\newtheorem*{theorem_prime}{Theorem 1'}
\theoremstyle{definition}
\newtheorem{defn}{Definition}
\newtheorem{ass}{Assumption}
\newtheorem*{question}{Central Question}
\theoremstyle{remark}
\newtheorem{cor}[thm]{Corollary}

\begin{document}

\title{The extraction of work from quantum coherence}  

\author{Kamil Korzekwa$^*$}
\affiliation{Department of Physics, Imperial College London, London SW7 2AZ, United Kingdom}
\author{Matteo Lostaglio$^*$}
\affiliation{Department of Physics, Imperial College London, London SW7 2AZ, United Kingdom}
\author{Jonathan Oppenheim}
\affiliation{Department of Physics, University College London, London WC1E  6BT, United Kingdom}
\author{David Jennings}
\address{Department of Physics, Imperial College London, London SW7 2AZ, United Kingdom}

\begin{abstract}
The interplay between quantum-mechanical properties, such as coherence, and classical notions, such as energy, is a subtle topic at the forefront of quantum thermodynamics. The traditional Carnot argument limits the conversion of heat to work; here we critically assess the problem of converting coherence to work. Through a careful account of all resources involved in the thermodynamic transformations within a fully quantum-mechanical treatment, we show that there exist thermal machines extracting work from coherence arbitrarily well. Such machines only need to act on individual copies of a state and can be reused. On the other hand, we show that for any thermal machine with finite resources not all the coherence of a state can be extracted as work. However, even bounded thermal machines can be reused infinitely many times in the process of work extraction from coherence.
\end{abstract}


\maketitle

\renewcommand{\thefootnote}{\fnsymbol{footnote}}
\footnotetext{These authors contributed equally to this work.}
\renewcommand{\thefootnote}{\arabic{footnote}} 
\setcounter{footnote}{0}
 
\section{Introduction}

\emph{Scientia potentia est}, knowledge is power, the Latin aphorism goes. This could not be more true in thermodynamics, where knowledge about the state of a system can be exploited to our advantage to extract work from it \cite{maxwell1872theory,szilard1929uber}. In quantum mechanics states of maximal knowledge are called pure states. A peculiar feature of the quantum world is that, due to the superposition principle, even for such states there are many questions that cannot be answered sharply. In thermodynamics we are especially interested in energetic considerations and so an odd place is taken by pure states that are a superposition of different energy states. This is because, despite the fact that we possess full knowledge about the system, our possibility of predicting the outcome of an energy measurement can be very limited. 

In standard quantum-mechanical considerations this is not a issue, because we can always reversibly transform a pure state into any other pure state by unitary dynamics. A basic task of thermodynamics, though, is the book-keeping of all energy flows from and out of the system, and there is no reversible transformation mapping a superposition of different energy states into an eigenstate while strictly conserving energy. Hence, we are left to wonder whether the ``scientia'' of having a pure state with quantum coherence can be converted into ``potentia'' of extracted work, while being limited by the law of energy conservation.

More precisely, we analyze work extraction from quantum coherence\footnote{Here, and in the rest of the paper, we use the term ``coherence'' in the sense of ``superposition of states belonging to different energy eigenspaces''.}, in the context of the theory of thermodynamics of individual quantum systems, currently under development \cite{janzing2000thermodynamic, horodecki2003reversible,dahlsten2011inadequacy, del2011thermodynamic,horodecki2013fundamental, brandao2013second, aberg2013truly, skrzypczyk2013extracting,  halpern2014unification,egloff2015measure,faist2015minimal,lostaglio2015stochastic, renes2014work,gemmer2015single}. The aim of the theory is to provide a suitable theoretical framework for our increasing ability to manipulate micro- and nanoscale systems \cite{collin2005verification, serreli2006molecular, toyabe2010experimental,alemany2010fluctuations, cheng2010bipedal}. Such general framework could also help approaching related questions, such as the role of quantum effects in biological systems \cite{lloyd2011quantum,lambert2013quantum,gauger2011sustained} and the link between thermodynamics and quantum information processing. Although this field has recently seen a great number of contributions, most of the previously mentioned works do not incorporate the possibility of processing a state in a superposition of energy eigenstates. It was only relatively recently that the role of coherence in thermodynamics has been looked at more closely \cite{scully2003extracting, brandao2011resource,rodriguez2013thermodynamics,lostaglio2015description,klimovsky2014heat, cwiklinski2014limitations,aberg2014catalytic,narasimhachar2014low,lostaglio2015quantum, uzdin2015quantum}.  

In this paper we first set the scene by presenting existing approaches to the problem of work extraction from the coherence of quantum systems. We argue that, within the regime of individually processed systems, the current approaches fail to account for all the resources used during the work extraction protocol. The typical assumption is to use a classical external field that experiences no back-reaction~\cite{skrzypczyk2014work,kammerlander2015quantum}. However, this does not allow for a full accounting of the thermodynamic cost of maintaining the field. Although this cost may be small in a single use, it has to be accounted for since the work gain will also be small. Hence we propose an alternative framework that aims for a careful book-keeping of resources.

In particular, we use the notion of a {\it thermal machine} \cite{masanes2014derivation},
a device of bounded resources that can be used to manipulate thermodynamical systems and perform tasks such as work extraction. Our thermal machine incorporates the use of an ancillary system carrying coherence (henceforth called \emph{reference} system), introduced into the context of coherence manipulation in thermodynamics in \cite{brandao2011resource}. It also includes a battery system where work can be stored, or transferred to the reference when necessary. A crucial question addressed in this work will be how to use the thermal machine in a repeatable way, i.e. without deteriorating it. We make use of an important result of Johan \AA berg, showing that reference systems can be used repeatedly to manipulate coherence \cite{aberg2014catalytic}. However the reference needs to be repumped for the machine to continue to operate. The work cost of repumping needs to be taken into account, which can mean that the thermal machine will not be able to extract all the available work from coherence. Nonetheless, we will find that we can come arbitrarily close, 
by choosing the amount of coherence resources carried by the reference system in the machine appropriately large. 
 
On the other hand, for any given thermal machine, we will prove that one can never extract all the available work. 
We will show that coherences of individual quantum systems can be exploited to enhance the performance of work extraction protocols (both in the average and the single-shot sense), but not to the extent that could be expected in the ``classical'' limit. Moreover, the work extraction protocol we provide does not deteriorate the thermal machine.

\section{Coherence and work}
 \label{sec:two}

\subsection{Setting the scene}

Let us start by introducing the framework that we will use throughout this paper and collect the core assumptions that our results rest upon. We want to study the allowed thermodynamic transformations by explicitly modelling any coherence resources being used. There are two ways in which coherence can enter the thermodynamics of the systems under consideration. This can happen either explicitly, by transferring it from an external system with quantum coherence (a trivial example being a swap operation between the system and an ancillary coherent state); or implicitly, by allowing operations that do not conserve energy (e.g., $\ket{0} \rightarrow \ket{+}$, where the Hamiltonian is given by $H_S=\ketbra{1}{1}$) or conserve it only on average (e.g., $\ket{1} \rightarrow (\ket{0}+\ket{2})/\sqrt{2}$, with $H_S=\ketbra{1}{1}+2\ketbra{2}{2}$). Therefore, we will only allow for those transformations that do not implicitly introduce coherence:
\begin{ass}[Allowed transformations]
\label{assumption1}
The set of allowed transformations is given by all (strictly) energy-preserving unitaries, i.e., unitaries that commute with the total free Hamiltonian of the system. The use of all ancillary systems should be \emph{explicitly} accounted for.\footnote{As we will see, thermal ancillas are the only ones that can be freely introduced without trivialising the problem of work extraction.}
\end{ass}
\noindent We will also take a closer look at an alternative approach in Sec.~\ref{sec:average} and explain why we find it not satisfactory for the aims of the present work. 

In this paper we focus on the task of work extraction from quantum systems with coherence. We do not aim here to settle the long-standing issue of what is an appropriate definition of work in quantum thermodynamics (see, e.g., \cite{frenzel2014reexamination,gallego2015defining}). For the scope of this paper we will assume, for the sake of simplicity, that the following holds for classical (incoherent) states:
\begin{ass}[Average work, incoherent states]
\label{assumption2}
Let $\rho_S$ be a quantum state of the system described by Hamiltonian $H_S$, with $\rho_S$ being incoherent in the energy eigenbasis. Then, in the presence of a heat bath at temperature $T$, an average amount of work $\langle W \rangle (\rho_S)$ equal to the change of free energy of a state can be extracted from it:
\begin{equation}
\label{eq:free_energy}
\langle W \rangle (\rho_S) = \Delta F(\rho_S) := F(\rho_S) - F(\gamma_S),
\end{equation}
where $F(\sigma) := \trace{}{\sigma H_S} - kT S(\sigma)$, $S(\cdot)$ is the von Neumann entropy and \mbox{$\gamma_S = e^{-H_S/kT} / Z_S$} is a thermal state with $Z_S$ being the partition function of the system.
\end{ass}
\noindent This formula, consistent with traditional thermodynamics, has been obtained using work extraction models that differ in details, but agree on the result \cite{aberg2013truly,skrzypczyk2014work}. For example, in Ref~\cite{aberg2013truly} the work extraction protocol is based on two elementary processes: level transformations (that change the eigenvalues of the system Hamiltonian $H_S$) and full thermalisation with respect to the current system Hamiltonian (through thermal contact with a bath at temperature $T$). Average work is then defined as the average change in energy during level transformations (the ``unitary'' steps) and, if initial and final Hamiltonian coincide, Eq. \eqref{eq:free_energy} is recovered. Here we focus on the problem of extending Eq.~\eqref{eq:free_energy} to quantum states with coherence. The results of the present paper apply to any definition of work satisfying Assumption \ref{assumption2}.

The problem of work extraction can also be studied in the so-called single-shot regime. This means that one is interested in single instances of the work extraction protocol, instead of average quantities. To explain this more precisely, let us refer again to the model introduced in Ref.~\cite{aberg2013truly} that we have summarised above. Extracted work can then be seen as a random variable, maximising the average of which yields Eq.~\eqref{eq:free_energy}. However, we may instead ask what is the maximum amount of deterministic (i.e., fluctuation-free) work that can be extracted during a single instance of the protocol, while allowing the failure probability $\epsilon$. In Ref~\cite{aberg2013truly} it was shown that for incoherent states this quantity is given by $F^\epsilon_0(\rho_S) - F(\gamma_S)$, where $F^\epsilon_0(\rho_S) = - kT \log Z_{\epsilon}$ is a single-shot free energy defined as follows. Given a subset $\Lambda$ of the indices $\{i\}$ labelling the energy levels of the system, define $Z(\Lambda) = \sum_{i \in \Lambda} e^{-\beta E_i}$, where $E_i$ are the eigenvalues of $H_S$. Then $Z_\epsilon = \min_{\Lambda} \{Z(\Lambda): \sum_{i\in \Lambda} p_i > 1-\epsilon\} $, where $p_i = \bra{E_i}\rho_S\ket{E_i}$. This result is also in agreement with other work extraction models based on thermal operations \cite{horodecki2013fundamental}. Hence, for the single-shot scenario we can use the following assumption:
\begin{ass}[Single-shot work, incoherent states]
\label{assumption3}
Let $\rho_S$ be a quantum state of the system described by Hamiltonian $H_S$, with $\rho_S$ being incoherent in the energy eigenbasis. Then in the presence of a heat bath at temperature $T$, using a single-shot protocol one can extract a sharp amount of work $W^\epsilon_{ss}$ with failure probability $\epsilon$:
\begin{equation}
\label{eq:ss_work}
W^\epsilon_{ss}(\rho_S) = \Delta F^\epsilon_0(\rho_S) := F^\epsilon_0(\rho_S) - F(\gamma_S).
\end{equation}
\end{ass}
\noindent Once again, our aim is to extend Eq.~\eqref{eq:ss_work} to quantum states with coherence and our results apply to any definition of single-shot work satisfying Assumption \ref{assumption3}. For the sake of brevity in the remaining of this paper we will only write ``extracting work equal to the free energy'', omitting ``in the presence of a heat bath at temperature $T$''; however, this is how our claims should be understood.

In thermodynamic considerations thermal Gibbs states are the only ancillary states that can be introduced without the need for careful accounting. In fact, one can show that using energy-preserving unitaries (in accordance with Assumption~\ref{assumption1}), a thermal state is the only one that can be introduced for free without allowing the production of every incoherent state~\cite{brandao2013second}. Clearly, if this was possible, then from Assumptions~\ref{assumption2}~and~\ref{assumption3} one could extract infinite amount of work, thus trivialising the theory. Hence, the most general thermodynamic transformations that can be performed without using extra resources are given by:
\begin{enumerate}
\item adding a bath system in a thermal state $\gamma_E$ with arbitrary Hamiltonian $H_E$ and fixed inverse temperature $\beta = 1/kT$,
\begin{equation}
\rho_S \mapsto \rho_S \otimes \gamma_E, \quad \gamma_E = e^{-\beta H_E}/\trace{}{e^{-\beta H_E}};
\end{equation}
\item performing any global unitary that conserves total energy, i.e., that commutes with the total free Hamiltonian of the system and baths, in accordance with Assumption~\ref{assumption1};
\item discarding any subsystem.
\end{enumerate}
The set of quantum maps acting on a system that arise from combining the transformations described above is known under the name of \emph{thermal operations} \cite{janzing2000thermodynamic, brandao2011resource}. 

\subsection{Work-locking}
\label{sec:worklocking}

The aim of this work is to begin with a system initially in a state with coherence $\rho_S$, and finish with a thermal state $\gamma_S$, while optimally increasing the free energy of a battery (storage) system. The initial and final battery states, $\rho_B$ and $\rho_B'$, should be incoherent, so that using Assumptions~\ref{assumption2}~and~\ref{assumption3} we can achieve the coherence to work conversion that we are looking for. Schematically:
\begin{equation}
\label{eq:aim}
\rho_S \otimes \rho_B\rightarrow \gamma_S \otimes \rho_B'.
\end{equation}
Without the use of an ancillary resource state the above transformation is given by a thermal operation. Note that thermal operations commute with the dephasing channel $\D$ \cite{brandao2011resource} that removes all coherence from a quantum state, 
\begin{equation*}
\D(\sigma) := \sum_i \trace{}{\Pi_i \sigma} \Pi_i,
\end{equation*}
where $\Pi_i$ are the projectors on the energy eigenspaces of the system under consideration. Hence, we get that if the transformation described by Eq.~\eqref{eq:aim} is possible, also the following one is:
\begin{equation*}
\D(\rho_S)\otimes \rho_B \rightarrow \gamma_S \otimes \rho_B'.
\end{equation*}
This implies $\Delta F(\rho_B) \leq \Delta F(\mathcal{D}(\rho_S))$, because $F$ is non-increasing under thermal operations. From Assumption~\ref{assumption2} we then have \mbox{$\langle W \rangle (\rho_S) \leq \langle W \rangle (\D(\rho_S))$}. A similar argument gives also $W^\epsilon_{\rm{ss}}(\rho_S)\leq W^\epsilon_{\rm{ss}}(\D(\rho_S))$; note that in both cases the bound is achievable because dephasing is a thermal operation. This phenomenon was observed before \cite{horodecki2013fundamental,skrzypczyk2013extracting} and was called ``work-locking'' in \cite{lostaglio2015description}. Work-locking highlights that, despite contributing to the free energy of the state, quantum coherence does not contribute to work extraction: it is ``locked''. It also shows, in agreement with \cite{aberg2014catalytic}, that the standard formula $\langle W \rangle (\rho_S)=\Delta F(\rho_S)$ applied to every state (also the ones with coherence), implicitly assumes the access to an external source of coherence. In this paper we revise the problem of extracting work from coherence, clarifying the role of this external source of coherence. To summarise
\begin{question}
To what extent can Eq.~\eqref{eq:free_energy} and Eq.~\eqref{eq:ss_work} be extended to arbitrary quantum states with coherence, while explicitly accounting for coherence resources (ancillary systems)?
\end{question}

\subsection{Different thermodynamic regimes}
\label{sec:unlocking}

With an increasing interest in the thermodynamics of non-equilibrium quantum systems, an important distinction to make is between ``single-shot'' statements, which are valid for every run of the protocol, and ``many-runs'' statements, valid in the case of a large number $M$ of runs. In the \emph{asymptotic regime} $M \rightarrow \infty$ one is focused on studying average quantities (like average extracted work), which is justified by the fact that the fluctuations around the average can be made negligible in the limit of a large number of runs of the protocol (which is often the situation of interest in the study of heat engines). On the other hand, although the expected amount of extracted work can be studied in a single-shot regime \cite{skrzypczyk2014work}, it potentially carries little information about the system at hand due to the large fluctuations of non-equilibrium thermodynamics \cite{aberg2013truly}. Instead, the focus in the single-shot regime is typically on probabilistic work extraction protocols that guarantee precise and sharp amount of work with a finite probability of success or some minimum amount of guaranteed work \cite{dahlsten2011inadequacy,horodecki2013fundamental,aberg2013truly, halpern2014unification}.
On top of this classification we can also differentiate between ``individual processing'' scenarios, in which a single (possibly nanoscale) system undergoes a thermodynamic process on its own; and ``collective'' scenarios, in which $N>1$ copies of a state are processed together (the $N\rightarrow \infty$ limit is considered in \cite{brandao2011resource}).

\begin{figure}[t!]\centering
\includegraphics[width=0.8\columnwidth]{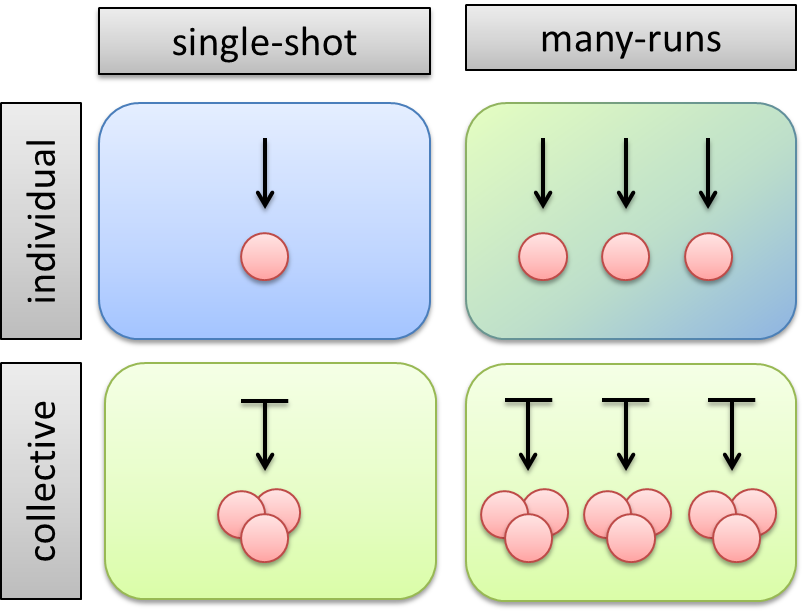}
\caption{\label{fig:singlemany} \textbf{Thermodynamic regimes}. Work extraction protocols can be investigated in different thermodynamic regimes. These can be classified by the number of systems that are processed at each run of the protocol (individual vs collective) and the number of times the protocol is repeated (single-shot vs many-runs). The green background indicates that in a given regime the maximal amount of work that can be extracted is consistent with traditional thermodynamics.}
\end{figure}

This classification of the thermodynamic regimes in which work extraction can be analysed is presented in Fig.~\ref{fig:singlemany}. The work-locking described in the previous section is a feature appearing in the regime of individually processed quantum systems. On the other hand, allowing for collective processing of the systems or for an ancillary quantum memory, one can ``unlock'' work from quantum coherence (see Appendix~A for a short description of this point). For arbitrary collective processes, one only needs to use
a sublinear amount of coherence in the reference system, meaning that its consumption does not contribute to the average work consumed or produced in the $N\rightarrow\infty$ limit~\cite{brandao2011resource}. In this limit coherence plays no role as, e.g., for $N$ identical qubits \mbox{$F(\rho^{\otimes N})\approx F(\D(\rho^{\otimes N}))$}, with a deficit scaling as $\log N/N$ \cite{lostaglio2015description}. Hence, in this paper we are interested in the thermodynamics of individual quantum systems ($N=1$, the upper half of Fig.~\ref{fig:singlemany}).

\subsection{Individual processing regime}

\subsubsection{Average energy conservation}
\label{sec:average}

In \cite{skrzypczyk2014work} sharp energy conservation, as expressed by the unitary dynamics commuting with the total Hamiltonian $H_{\rm tot}$, was replaced with the condition that such dynamics only keeps the first moment $\<H_{\rm tot}\>$ constant. Under this weaker condition it was shown that an amount of energy equal to the free energy difference $\Delta F(\rho_S)$ can be extracted on average from a system in an arbitrary quantum state $\rho_S$.

The elegance and appeal of this is that it recovers a clear thermodynamic meaning for the free energy of an individual quantum system. However, several problematic issues can be raised. Firstly, if one is interested in analysing the class of allowed quantum operations, then in the average-energy scenario this set depends on the particular state one is processing, which is conceptually less appealing and technically problematic from a resource-theoretic perspective.

Secondly, restricting energy considerations to the first moment analysis can hide arbitrarily large energy fluctuations described by higher moments, that are not explicitly modelled, but may be highly relevant. To see this consider a unitary $U_{\mathrm{\rm ave}}$ mapping a state \mbox{$\ket{\psi_{02}}:=(\ket{0}+\ket{2})/\sqrt{2}$} to $\ket{1}$, which preserves energy on average (here $\ket{n}$ is the energy eigenstate corresponding to energy $n$). Since microscopically all processes are ultimately energy-conserving, $U_{\mathrm{ave}}$ must be realized through a joint energy-preserving unitary $U$ involving $\ket{\psi_{02}}$ and some ancillary state $\rho_A$, e.g., the state of the battery,
\begin{equation*}
U(\ketbra{\psi_{02}}{\psi_{02}}\otimes\rho_A)U^{\dagger}=\ketbra{1}{1}\otimes\rho_A'.
\end{equation*}
In any such process the energy fluctuations of the ancillary system must increase. Specifically, denoting by $H$ the Shannon entropy of the outcomes of an energy measurement one gets a strict inequality: $H(\rho_A') > H(\rho_A)$ (see Appendix B for details).
As an example consider the ancillary system prepared in the energy eigenstate $\ket{m}$, so that an energy measurement would give a sharp outcome. Then, while the system is transformed from $\ket{\psi_{02}}$ into $\ket{1}$, the ancilla must be transformed into a superposition of energy eigenstates $(\ket{m+1} + \ket{m-1})/\sqrt{2}$. Hence an energy measurement would show fluctuations in the final state of the ancilla. It is important to note that the protocol that extracts work from coherence within the framework of average energy conservation \cite{skrzypczyk2014work}, necessarily creates such extra fluctuations, however these are not explicitly modelled within the formalism used. As we will see it is exactly due to these fluctuations that our protocols require work to be invested in restoring the ancillary state.

Finally, as the fluctuations created by operations that conserve energy only on average remain outside the formalism, one cannot properly account for the fluctuations in the extracted work outside the asymptotic regime.

\subsubsection{Repeatable use of coherence resources}

As already mentioned in Sec. \ref{sec:worklocking}, in the presence of energy conservation and without additional coherence resources, work-locking prevents us from extracting work from the coherence of individual quantum systems. One could then stay within the framework of strict energy conservation, but allow for the use of an extra source of coherence. We refer to this extra system as the \emph{reference}. 

At one extreme one could allow for the use of an infinite source of coherence (an ``unbounded'' reference frame \cite{aharonovsusskind,bartlett2007reference})\footnote{Not to be confused with a reference described by a Hamiltonian unbounded from below, which is unphysical.}, that entirely negates the constraints and experiences no back-reaction from its use on the quantum system. As suggested in \cite{lostaglio2015description}, in such case we should be able to extract all the work from coherence. However, one might worry that this involves the accounting ``$\infty - c = \infty$'', with $c$ being some finite resource consumed from an infinitely large reference system. Indeed, the use of such an unbounded reference allows us to simulate the operations from the previous section (conserving energy only on average) \cite{bartlett2007reference}, and hide the arising extra fluctuations in the infinitely big reference system. This semiclassical treatment is typical for many standard approaches that assume the existence of a classical field experiencing no back-reaction from the system \cite{skrzypczyk2014work,kammerlander2015quantum} and works well in many circumstances. However we are interested in the regime in which the thermal machine itself may be a microscopic quantum system. Hence, it seems more reasonable to firstly consider the reference as a quantum system with finite coherence resources -- a ``bounded'' reference frame -- and only then study the limit of an unbounded reference (recent works in this spirit and the discussion of semiclassical approaches can be found, e.g., in \cite{klimovsky2013work, klimovsky2014heat,frenzel2014reexamination}).

\begin{defn}[Reference]\label{def:reference}
We consider a reference (or coherence reservoir) given by an infinite-dimensional ladder system described by Hamiltonian \mbox{$H_R = \sum_{n=0}^\infty n \ketbra{n}{n}$}. We characterise the state $\rho_R$ of the reference through two numbers, $(\langle \bar{\Delta} \rangle, M)$. The first parameter, $\langle \bar{\Delta} \rangle$, measures the coherence properties of the reference and is given by 
\begin{equation}
\langle \bar{\Delta} \rangle = \trace{}{\rho_R \bar{\Delta}}, \quad \bar{\Delta} = (\Delta + \Delta^\dag)/2,
\end{equation}
where $\Delta$ is the shift operator $\Delta = \sum_{n=0}^\infty \ketbra{n+1}{n}$. We have that $\langle \bar{\Delta} \rangle < 1$ and the limit case $\langle \bar{\Delta} \rangle =1$ is called \emph{unbounded} or classical reference. The second parameter, $M$, describes the lowest occupied energy state, $M= \min \{n: \bra{n}\rho_R \ket{n} > 0\}$. 
\end{defn}
\noindent Examples of a sequence of references that come arbitrarily close to a classical one are uniform superpositions of $L$ energy states when $L \rightarrow \infty$ or coherent states with arbitrarily large amplitude. The use of $\langle \bar{\Delta} \rangle$ and $M$ as relevant quality parameters will soon become clear.

Results from the field of quantum reference frames \cite{aharonovsusskind,bartlett2006degradation, bartlett2007reference, white2009consumption, ahmadi2010dynamics} suggest that the back-reaction experienced by the reference will necessarily deteriorate it and consume the resources. 
However, if the usefulness of the reference or field is continually degraded during the work extraction process, we cannot claim that we are presenting a protocol performing work extraction from the state alone, as extra resources are consumed. Similar problems arise if free energy is continually taken away from the reference. 

In this paper we propose the following approach. We allow for the use of additional coherence resources as part of our thermal machine, but demand that they are used \emph{repeatably} in the following sense: the performance of our reference-assisted protocol, while operating individually on the $n$-th copy of the system, must be the same as while operating on the \mbox{$(n+1)$-th} copy, for all $n \in \mathbb{N}$. In other words, repeatability means that the reference's ability to perform the protocol never degrades, but crucially its state is allowed to change. Essentially this means that despite that the free energy of the reference can fluctuate and its coherence properties change, it can be used indefinitely to repeat the same protocol. To design such a protocol we employ the recent surprising result of \cite{aberg2014catalytic} that shows how a coherence resource can be used repeatably to lift the symmetry constraints 
imposed by energy conservation.\footnote{The work \cite{aberg2014catalytic} actually uses the word ``catalysis'', but we prefer to use the word repeatability/repeatable to avoid suggesting that there is no change in the state of the reference. Recall that traditionally a catalyst is a system in a state $\chi$ that enables $\rho\otimes \chi \rightarrow \sigma \otimes \chi$, despite $\rho \rightarrow \sigma$ being impossible (see, e.g., \cite{horodecki2009quantum, brandao2013second}). Repeatability, on the other hand, only requires the auxiliary system to be as useful at the end as it was at the beginning, while its state may change.} However, as we will see the protocol in \cite{aberg2014catalytic} requires continuous injection of energy into the reference (we do not allow the Hamiltonian of the reference to be unbounded from below, as in \cite{malabarba2015clock}). Hence, it is not immediately obvious that net thermodynamic work can be extracted from coherence. 

In what follows we introduce a general protocol that processes quantum systems individually and allow us to extract work from their coherence. We then focus on two variations of it. The first one can come arbitrarily close to extracting all the coherence as average work with arbitrarily small failure probability, provided we make the coherence resources of the reference system in the thermal machine large enough. However, if one does not have access to arbitrarily large coherence resources, this variation of the protocol does not guarantee perfect repeatability. Therefore, we examine a second variation that is perfectly repeatable even for bounded references.
We show then that the performance of work extraction in both the single-shot and asymptotic regimes is enhanced only if the quality of the reference (defined further in the text) is above a certain threshold.

\section{The protocol}
\label{sec:protocol2}

We analyze work extraction from pure qubit states with coherence,
\begin{equation}
\label{eq:psi}
\ket{\psi}=\sqrt{1-p}\ket{0}+ \sqrt{p}e^{-i \varphi}\ket{1}, \quad p \in (0,1).
\end{equation}
Without loss of generality we can set $H_S = \ketbra{1}{1}$ and $\varphi =0$ (rotations about the $z$ axis of the Bloch sphere conserve energy). Our aim is to unlock work from coherence through the repeatable use of a thermal machine containing a reference, while processing each copy of $|\psi\>$ individually. 
In Table \ref{table:protocols} the results we obtain within this framework are schematically compared with the ones obtained within the frameworks presented in the previous section for the paradigmatic example of a qubit in a ``coherent Gibbs state'' $\ket{\gamma}$ given by:
\begin{equation}
\label{eq:ketgamma}
\ket{\gamma}=\sqrt{1-r}\ket{0}+\sqrt{r}\ket{1},
\end{equation}
with $(1-r,r)$ being the thermal distribution for the system, so that $\D(\ketbra{\gamma}{\gamma})=\gamma_S$.

The extraction of non-zero work from $\ket{\psi}$ then requires a thermal machine containing a reference state $\rho_R$ and implementing an energy-conserving unitary $V$:
\begin{equation*} 
\rho_{SR}'=V(\ketbra{\psi}{\psi}\otimes \rho_R)V^{\dagger},
\end{equation*}
satisfying the following: 
\begin{enumerate}
\item The system is pre-processed to a new state \mbox{$\rho_S' = \trace{R}{\rho_{SR}'}$} that allows for better work extraction than from the initial state $\ket{\psi}$.
\item The final reference state $\rho'_R=\trace{S}{\rho'_{SR}}$ can be processed into a state $\rho''_R$ (perhaps using some of the extracted work) in such a way that the repeatability requirement is satisfied.
\item No collective operations, at any stage of the protocol, are allowed on multiple copies of $\ket{\psi}$ and no quantum memory (in the sense of Appendix A) is used.
\end{enumerate}
\begin{table}
\renewcommand{\arraystretch}{1.2}
\begin{tabular}{ c | c  c |}
 \cline{2-3}                  
   & Single-shot & \multicolumn{1}{|c|} {Asymptotic} \\ 
 \hline  \hline 
\multicolumn{1}{|c|} {Average energy} & $\< W\>=\Delta F$ \cite{skrzypczyk2014work}  & \multicolumn{1}{|c|} {\multirow{2}{*}{$\< W\>=\Delta F$ \cite{skrzypczyk2014work}}} \\ 
\multicolumn{1}{|c|} {conservation} & Large fluctuations \cite{aberg2013truly}& \multicolumn{1}{|c|} {}  \\
\hline 
\multicolumn{1}{|c|}{Strict energy} & \multicolumn{2}{ |c| }{\multirow{2}{*}{$\epsilon(\ket{\gamma})=\epsilon(\gamma_S)$, $\< W\>=0$ \cite{horodecki2013fundamental,skrzypczyk2013extracting,lostaglio2015description}}}\\
\multicolumn{1}{ |c|  }{conservation} & \multicolumn{2}{ |c| }{} \\
\hline 
\multicolumn{1}{|c|}{Strict energy} & $\epsilon(\ket{\gamma})=0$ & \multicolumn{1}{|c|} {$\< W\>=\Delta F$} \\
\multicolumn{1}{|c|}{conservation w/}&unbounded reference  &\multicolumn{1}{|c|} {unbounded reference}\\
\multicolumn{1}{|c|}{resource used}& $\epsilon(\ket{\gamma})<\epsilon(\gamma_S)$ & \multicolumn{1}{|c|} {$\< W\><\Delta F$}\\
\multicolumn{1}{|c|}{repeatably}& bounded reference &\multicolumn{1}{|c|} {bounded reference}\\
 \hline  
\end{tabular}
\caption{\label{table:protocols} \textbf{Individual processing protocols extracting work from $\ket{\gamma}$.} $\< W\>$ denotes the average work that can be extracted from the coherent thermal state $|\gamma\>$ and $\epsilon$ denotes the error probability of a single-shot work extraction from a given state. A thermal state of the system is denoted by $\gamma_S$. Note that under operations strictly conserving energy, no work extraction protocol on $\ket{\gamma}$ can outperform a work extraction protocol on $\gamma_S$, as the two states are indistinguishable.}

\end{table}

\subsection{The explicit work-extraction protocol}
\label{sec:protocol}

A protocol satisfying the introduced requirements consists of the following steps (see Fig.~\ref{fig:protocol}):
\begin{enumerate}
\item\label{step1} \textbf{Pre-processing.} The system $\ket{\psi}$ interacts through an energy-preserving unitary $V(U)$ with the reference $\rho_{R}$. The unitary acting on the joint system $SR$ is chosen as in \cite{aberg2014catalytic} to be:
\begin{equation}
\label{abergdynamics}
V(U)=\ketbra{0}{0}\otimes\ketbra{0}{0}+\sum_{l=1}^{\infty} V_l(U),
\end{equation}
with
\begin{equation*}
V_l(U)=\sum_{n,m=0}^{1} \matrixel{n}{U}{m} \ketbra{n}{m}\otimes\ketbra{l-n}{l-m}. 
\end{equation*}

We choose
\begin{equation*}
U=\left(\begin{array}{cc}
\sqrt{p}&-\sqrt{1-p}\\
\sqrt{1-p}&\sqrt{p}
\end{array}\right),
\end{equation*}
so that $U$ rotates the qubit system from $\ket{\psi}$ to $\ket{1}$.\footnote{This interaction corresponds to a modified Jaynes-Cummings model (with excitation-dependent coupling strengths). However, it can also be approximately realized within the standard Jaynes-Cummings model using a reference in a coherent state $|\alpha\>$, with $|\alpha|$ large enough (for details see Supplementary Material Sec.~V in \cite{aberg2014catalytic}).}

\item\label{step2} \textbf{Work extraction. }The system is now in a state $\rho_S'$ and, due to work-locking (see Sec. \ref{sec:worklocking}), is indistinguishable from its dephased version in any work extraction protocol. So without loss of generality we can use the dephased version
\begin{equation}
\label{eq:dephased}
\mathcal{D}(\rho_S') = (1-q) \ketbra{0}{0} + q \ketbra{1}{1}.
\end{equation}
Now, depending on the considered regime, single-shot or average (asymptotic) work can be extracted from the dephased state and stored in the thermal machine.

\begin{figure}[t!]
\centering
\includegraphics[width=0.95\columnwidth]{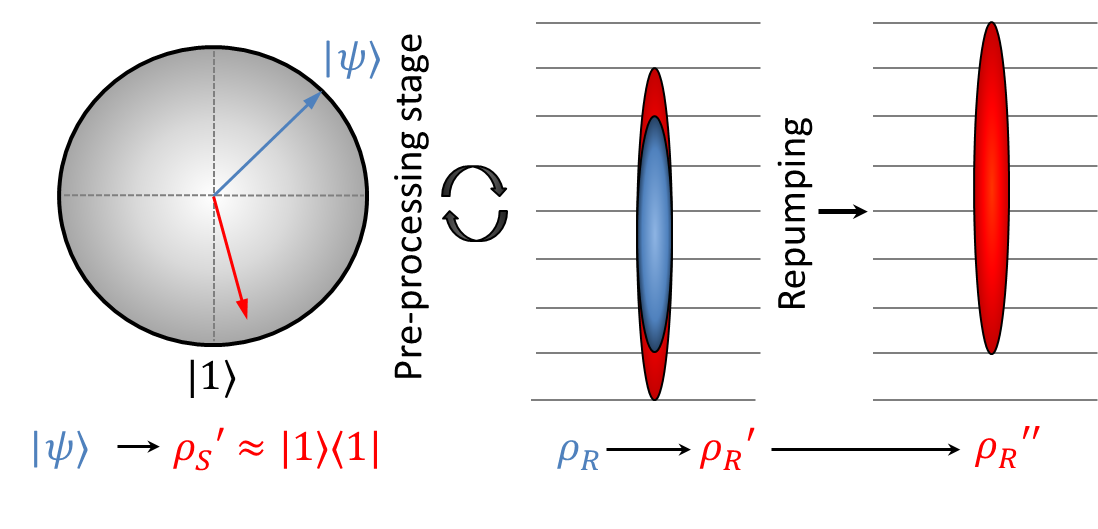}
\caption{\label{fig:protocol}  \textbf{The basic protocol}.
The evolution of the system from the initial state $\ket{\psi}$ to the final state is depicted on the Bloch ball, in blue and red respectively. The evolution of the reference from the initial state (smaller blue blob) throughout the protocol (red blobs) is depicted on the energy level ladder.}
\end{figure}

\item\label{step3} \textbf{Repumping. }The back-reaction changes the state of the reference into $\rho'_R$. Using part of the extracted work (stored in the battery during the previous step) we can repump the reference to shift it up:
\begin{equation}
\label{eq:repumping}
\rho'_R \rightarrow \rho_R'':=\Delta \rho'_R\Delta^{\dagger},
\end{equation}
with the details of how to perform such operation given in Appendix C. We will describe how often we perform the repumping while analysing different variations of the protocol.
\item We can repeat the protocol using $\rho''_R$ and a fresh copy of $\ket{\psi}$.
\end{enumerate}

\subsection{Performance}

During the pre-processing stage the joint unitary $V(U)$ approximately induces $U$ on the system:
\begin{equation*}
\rho_S' \approx U(\ketbra{\psi}{\psi})U^{\dagger} = \ketbra{1}{1}.
\end{equation*}
The degree to which the above equation holds depends on the \emph{quality} of the reference as defined in \cite{aberg2014catalytic}. In particular, the system final occupation in the excited state $q=\matrixel{1}{\rho_S'}{1}$ is given by
\begin{equation}
\label{eq:q}
q=1-2p(1-p)(1-\langle\bar{\Delta}\rangle)-(1-p)^2R_{00},
\end{equation}
where $R_{00} = \bra{0}\rho_R \ket{0}$ and $\bar{\Delta}$ is the quality parameter from Definition~\ref{def:reference}. From Eq.~\eqref{eq:q} it is easy to see that $q \rightarrow 1$ when $R_{00} \rightarrow 0$ (i.e., when the reference quality parameter $M>0$) and $\langle\bar{\Delta}\rangle \rightarrow 1$. Therefore, $R_{00}$ and $\langle\bar{\Delta}\rangle$ are operationally well-defined quality parameters of the reference, because they directly measure the ability of the reference to induce the unitary $U$ that we want to perform on the qubit. At the same time the reference undergoes a back-reaction induced by the joint unitary $V(U)$. This is described by the following Kraus operators:
\begin{subequations}
\begin{eqnarray}
A_0&=&(1-\sqrt{p})\sqrt{1-p}\ketbra{0}{0}+\sqrt{p(1-p)}(\iden-\Delta),\quad\quad\label{eq:kraus1}\\
A_1&=&p\iden+(1-p)\Delta^{\dagger}, \label{eq:kraus2} 
\end{eqnarray}
\end{subequations}
so that the reference state after performing the pre-processing stage is given by:
\begin{equation}
\label{eq:back}
\rho_R'=A_0\rho_RA_0^{\dagger}+A_1\rho_RA_1^{\dagger}.
\end{equation}

From Eqs.~\eqref{eq:dephased} and \eqref{eq:q} the only two parameters relevant for work extraction are the reference population in the ground-state, $R_{00}$, and the parameter $\langle \bar{\Delta} \rangle$. Using the Kraus operators specified by Eqs.~\eqref{eq:kraus1}-\eqref{eq:kraus2}, the change in $\langle \bar{\Delta} \rangle$ during the pre-processing stage (i.e., the difference between the final and initial value of $\langle \bar{\Delta} \rangle$) can be computed and is found to be:
\begin{equation}
\label{eq:deltadelta}
\delta \langle \bar{\Delta} \rangle = (1-p)\left[(1+\sqrt{p} -2p)\mathrm{Re} (R_{01})-\sqrt{p}R_{00}\right],
\end{equation}
where as before $R_{ij} = \bra{i}\rho_R\ket{j}$. A sufficient condition for $\langle\bar{\Delta}\rangle$ to stay constant is $R_{00}=0$, i.e., $M>0$. Therefore, if the initial state satisfies $R_{00}=0$ exactly, performing the pre-processing stage does not change $\langle\bar{\Delta}\rangle$, as noted in \cite{aberg2014catalytic}. We require step~(\ref{step3}) of the protocol to ensure that $\bra{0}\rho_R''\ket{0} = 0$. If this is the case, at the end of the protocol we are left with the reference described by the same quality parameters $R_{00}$ and $\langle\bar{\Delta}\rangle$ as at the beginning, and the reference $\rho_R''$ is as good as $\rho_R$ within the protocol. 

Finally, because the state of the reference changes, its free energy can fluctuate. However,     
notice firstly that the reference has Hamiltonian bounded from below, so for fixed average energy it has a finite amount of free energy. Secondly, repeatability requires that the reference can be used an arbitrary number of times and the performance of the protocol never changes. It is then easy to see that on average the free energy of the reference cannot be extracted as work, as this would be incompatible with repeatability. It can be shown that in the worst-case scenario the free energy change in the reference fluctuates around zero (see Appendix D for more details). Moreover, these fluctuations vanish in the limit of an unbounded reference $\langle \bar{\Delta} \rangle \rightarrow 1$, as then the entropy of the reference stays constant, while its average energy increases. Therefore its free energy must increase at every step of the protocol.

\section{Fundamental limitations of coherence to work conversion}

How well does the above approach do in terms of work extraction? Here, we first emphasize the limitations on work extraction from coherence that arise due to the reference being bounded, i.e., when we have access to limited coherence resources. More precisely, we will explain why the use of a bounded reference does not allow one to extract from a state with coherence the average amount of work equal to the free energy difference.

However, we will then also show that one can construct a series of bounded reference states that come arbitrarily close to extract the free energy $\Delta F(|\psi\>)$, with protocols arbitrarily close to perfect repeatability\footnote{A similar result appears in \cite{aberg2014catalytic}, however it was based on using a reference system described by a doubly-infinite ladder Hamiltonian. This left open the question if this limit is achievable by a system with a physically realisable Hamiltonian.}. Thus, we will prove that in the limit of an unbounded reference all coherence can be converted into work in a repeatable way. The limit case does not generate any entropy in the reference system and, being a reversible transformation, is optimal.

\subsection{Limitations of bounded thermal machines}

In order to illustrate the limitations arising from using a bounded reference we will consider a particular model of work extraction from coherence described in Refs.~\cite{skrzypczyk2014work,kammerlander2015quantum}. It has been proved there that the free energy difference $\Delta F(\rho_S)$ can be extracted from a system $\rho_S$ as work if one allows for the use of operations that conserve the energy only on average. Let us briefly recall the protocol used to achieve this. It is composed of two stages; first, given a state $\rho_S$, work is extracted from coherence. The resultant state is given by $\mathcal{D}(\rho_S)$ (recall that $\mathcal{D}$ denotes the dephasing operation in the energy eigenbasis). Second, work is extracted from the incoherent state $\mathcal{D}(\rho_S)$. In accordance with Assumption~\ref{assumption2}, the latter process extracts $\Delta F(\mathcal{D}(\rho_S))$. Hence, the extraction of the full free energy $\Delta F(\rho_S)$ from a state $\rho_S$ is equivalent to the possibility of extracting $F(\rho_S)-F(\D(\rho_S))$ from the coherence of $\rho_S$. Notice that this quantity coincides with $kT A(\rho_S)$, where \mbox{$A(\rho_S)=S(\rho_S||\D(\rho_S))$} is a known measure of quantum coherence \cite{gour2009measuring}, and it quantifies the amount of free energy stored in coherence \cite{lostaglio2015description}. Hence, the amount of work that needs to be extracted on average from the coherence of a quantum state to achieve the free energy extraction limit for arbitrary quantum states is
\begin{equation}
\label{eq:cohbound}
W_{\rm{coh}}(\rho_S) = kT A(\rho_S).
\end{equation}

Without loss of generality we can write any state $\rho_S$ as \mbox{$\sum_n p_n \ketbra{\psi_n}{\psi_n}$} with \mbox{$p_{n+1}\leq p_n$}. Let us also denote the Hamiltonian of the system by \mbox{$H_S = \sum_n E_n \ketbra{E_n}{E_n}$}. In the protocol that allows to extract work from coherence in Ref.~\cite{skrzypczyk2014work}, the system $\rho_S$ interacts with a weight system in a gravitational field via the unitary
\begin{equation}
\label{skrunitary}
U_{\mathrm{ave}}= \sum_n \ketbra{E_n}{\psi_n} \otimes \Gamma_{\varepsilon_n},
\end{equation}
where $\Gamma_{\varepsilon_n}$ is the shift operator on the weight system that shifts it in energy by $\varepsilon_n = \bra{\psi_n}H_S\ket{\psi_n} - E_n$. 

As $U_{\mathrm{ave}}$ does not strictly conserve energy, by Assumption~\ref{assumption1} it is not a free thermodynamic operation. One can instead ask if it can be achieved by an energy-preserving unitary $V(U_{\rm{ave}})$ on a larger system that exploits some ancillary system $\rho_A$ (that includes a battery). In other words, we are looking for the energy-preserving unitary $V(U_{\rm{ave}})$ such that
\begin{equation}
\label{eq:simulation}
\mathcal{E}(\rho_S):=\trace{A}{V(U_{\rm{ave}})(\rho_S \otimes \rho_A)V(U_{\rm{ave}})^\dag}=U_{\mathrm{ave}}\, \rho_S \, U_{\mathrm{ave}}^\dag.
\end{equation}
It is easy to show that due to the imposed constraints, the ancillary system $\rho_A$ must carry quantum coherence. In fact, if $\rho_A$ were incoherent, then the left-hand side of Eq.~\eqref{eq:simulation} would be a time-translation covariant quantum map (meaning that $[\mathcal{E},\D]=0$) \cite{lostaglio2015quantum}, whereas the right-hand side is not. 

Now the crucial point is that Eq.~\eqref{eq:simulation} cannot hold exactly unless $\rho_A$ contains an unbounded reference. If $\rho_A$ is bounded, then the reduced evolution of $\rho_S$ is not exactly unitary and not all the energy change can be identified with work. To prove this, one can compute the von Neumann entropy of both sides of Eq.~\eqref{eq:simulation} and notice that the mutual information $I(\rho_{SA})=0$. Note, however, that the only way this is possible is if $V(U_{\rm{ave}}) = V_1 \otimes V_2$, and further we would need $V_1 = U_{\mathrm{ave}}$. But this is not possible, because from the fact that $V(U_{\rm{ave}})$ is strictly energy-preserving we can prove that $V_1$ and $V_2$ must both be as well. Hence, the right-hand side of Eq.~\eqref{eq:simulation} cannot be a unitary if $\rho_A$ contains only a bounded reference frame. In fact, Eq.~\eqref{eq:simulation} can only hold as a limit case of using a larger and larger coherence resource. In summary, Assumptions \ref{assumption1}-\ref{assumption3} together with identification of work with energy change during unitary processes, imply that without an unbounded reference the work extraction protocol from Refs.~\cite{skrzypczyk2014work,kammerlander2015quantum} cannot extract an amount of work equal to $\Delta F(\rho_S)$ from a state with coherence.

The strength of this point is that we do not even need to require the repeatability of the protocol using the same ancillary system. In fact, the argument is rather general. In order to extract all the free energy from a state $\rho_S$ one needs to transform it into a thermal state. This cannot be achieved by only changing the energy spectrum of $H_S$, but also requires the rotation of the energy eigenbasis, so that the system is incoherent at the end of the transformation. This can be performed perfectly only with the aid of an unbounded reference frame, because it involves unitaries that do not strictly conserve energy.\footnote{A useful point of view is also given by the theory of quantum reference frames and recovery maps \cite{bartlett2007reference,bartlett2009quantum}.}

\subsection{Extracting work arbitrarily close to the free energy difference}
\label{sec:classical}

A key fact about the Carnot efficiency is that, despite being achieved only by ideal heat engines that do not actually exist in Nature, we can get arbitrarily close to it through a sequence of real engines. In a similar spirit, we now construct a sequence of bounded thermal machines getting arbitrarily close to the coherence to work conversion limit set by Eq.~\eqref{eq:cohbound}. The main result of this Section can be summarised in a non-technical way as follows:
\begin{numbered_theorem}
\label{theorem2}
There exists a sequence of bounded thermal machines approaching the ideal coherence to work conversion of Eq.~\eqref{eq:cohbound} with arbitrarily high probability of success and with an arbitrarily small change in the quality parameters. The limit case is reversible.
\end{numbered_theorem}
\noindent As an immediate consequence of the fact that the limit case is reversible we have:
\begin{cor}
Eq.~\eqref{eq:cohbound} provides the ultimate limit of coherence to average work conversion.
\end{cor}

In the remaining part of this section we give more details about the result above, first of all specifying the technical claim and then the main steps of the proof (the details of calculations can be found in Appendix E). We consider a sequence of reference states $\rho_R$ that approach a classical reference. Consider an arbitrary reference state $\rho_R$. We will describe it by two parameters $(\delt,M)$ according to Definition~\ref{def:reference}.

We will now show how to perform the protocol described in the previous section to extract from any pure state $\ket{\psi}$ an amount of work per copy arbitrarily close to the free energy difference $\Delta F(\ket{\psi})$, while succeeding with arbitrarily high probability and changing the quality of the reference only by a negligible amount. For simplicity, define $f(x)=-x - kT h_2(x)$, where $h_2(\cdot)$ denotes the binary entropy. Theorem~1 can be now made technically precise as follows:
\begin{theorem_prime}
\label{theorem3}
Let $\rho_R$ be an arbitrary reference state described by $(\delt,M)$. In the presence of a thermal bath at temperature $T$ and if $M$ is large enough, there exists a protocol individually extracting from $M$ copies of $\ket{\psi}$ [given by Eq.~\eqref{eq:psi}] an average amount of work $M \langle W \rangle$, with
\begin{equation*}
\langle W \rangle \geq \Delta F(\ket{\psi}) - f(2p(1-p)(1-\delt)) - O(M^{-\frac{1}{3}}).
\end{equation*}
The probability of success $p_{succ}$ of the protocol is 
\begin{equation*}
p_{\rm{succ}} \gtrsim [1- 2p(1-p) (1-\delt)]^M,
\end{equation*}
and it changes the quality parameters of the reference as follows:
\begin{equation*}
\delta M = 0, \quad \delta \delt \leq 2\sqrt{1-p_{\rm succ}}.
\end{equation*}
\end{theorem_prime}
\noindent Before presenting the proof of this theorem, let us first comment on its scope. Note that the same result holds when a reference $(\delt,M)$ is used a number of times $M'<M$, as long as $M' \gg 1$ (this will be clear from the proof). In the case in which $M'>M$, we can apply the Theorem every $M$ uses of the reference. The changes in the quality parameters will eventually sum up, but the Theorem gives a bound on them. Also, it will be clear from the proof that the failure of the protocol implies a destruction of the coherence properties of the reference.

We now prove the Theorem by constructing an explicit variation of protocol introduced in Sec.~\ref{sec:protocol} and showing that it performs as stated in the theorem. Recall that by an energy conserving unitary we can rotate $\ket{\psi}$ around the $z$ axis of the Bloch sphere. Hence, without loss of generality, we can set $\varphi=0$ in Eq.~\eqref{eq:psi}. We then perform steps (\ref{step1}) and (\ref{step2}) of the protocol described in Sec.~\ref{sec:protocol} $M$ times, i.e., individually processing each of $M$ copies of $\ket{\psi}$ using a reference $\rho_R$ described by $(\delt,M)$. The choice of $\rho_R$ ensures that during this process the reference state will have no population in the ground state, and so $\<\bar{\Delta}\>$ will stay constant. Then, the final state of the reference is described with probability $p_1^M$ by $\rho_{R,1} = A^M_1 \rho_R A^{\dag M}_1$, where \mbox{$p_1=(1-2p(1-p)(1-\delt))$}. Notice that having access to a reference described by parameters $\delt$ and $M$ such that $M(1-\delt)\rightarrow 0$, the probability $p_1^M$ can be made arbitrarily close to $1$. This happens because by taking $\delt$ close enough to 1, we can get arbitrarily close to unitary evolution of a system state $\ket{\psi}$ to a pure, incoherent state $\ket{1}$. As the final state of the system is almost pure, the final joint state of the system and the reference factorizes and an arbitrarily small amount of entropy is generated (the back-reaction on the reference is given by the Kraus operator $A_1$ alone). 

Next, we repump the reference $\bar{M}+s\sigma_M^{4/3}$ times, where $\bar{M}=M(1-p)$, \mbox{$\sigma_M=\sqrt{Mp(1-p)}$} and $s>0$. This guarantees that the reference has arbitrarily small population in states $\{\ket{0}\dots\ket{M}\}$, so that by performing a measurement we can project the reference to a state $\rho''_R$ with support on the subspace spanned by $\{\ket{i}\}_{i>M}$ with arbitrarily high probability ($s$ fixes the confidence level, see Appendix E for details). More precisely, after repeating steps (\ref{step1}) and (\ref{step2}) of the protocol $M$ times and repumping as explained above, the reference is described by a state $\rho_R''$ with probability
\begin{equation}
p_{\rm{succ}} \geq [1- 2p(1-p) (1-\delt)]^M \mathbb{E}_s(M^{1/6}),
\end{equation}
where $\mathbb{E}_s(x)={\rm erf}(sx/\sqrt{2})$ and $\rm{erf}$ denotes the error function. The final state is given by $\rho''_R$, described by $(\delt'',M'')$, where $M''=M$ and \mbox{$\delta \delt :=\delt'' -\delt$} is bounded as follows
\begin{equation}
\label{eq:fundamental}
\delta \delt \leq 2\sqrt{1-p_{{\rm succ}}} \leq 2\sqrt{1-p_1^M\mathbb{E}_s(M^{1/6})}.
\end{equation}
Notice that by taking $s$ large enough (but finite) we can make the factor $\mathbb{E}_s(M^{1/6})$ in the previous two equations arbitrarily close to $1$, $\mathbb{E}_s(M^{1/6}) \approx 1$. 
In the appropriately chosen limit $\delt\rightarrow 1$ and $M\rightarrow\infty$ the quality parameters of the reference state are then unchanged with probability 1. Let us also note that the cost $W_E$ of the measurement described above is bounded by $kT h_2(p_{\rm{succ}})$ (see Appendix~E). 

We have just shown that following the procedure above we can guarantee repeatability with arbitrary confidence level. Hence, we now proceed to proving that it also allows for extracting an average amount of work per system arbitrarily close to the free energy difference $\Delta F(\ket{\psi})$. To see this, note that after repeating the protocol on $M$ copies of $\ket{\psi}$ we are left with $M$ copies of a state $\D(\rho_S ')$ from Eq.~\eqref{eq:dephased} with $q$ given by Eq.~\eqref{eq:q}, \mbox{$q=1-2p(1-p)(1-\delt)$}. This state is diagonal in the energy eigenbasis and the average work $\langle \tilde{W} \rangle$ extracted from it is given by $\Delta F(\mathcal{D}(\rho_S'))$:
\begin{eqnarray}
\nonumber
\langle \tilde{W} \rangle &=& 1 + kT \log Z - 2p(1-p)(1-\delt)\\
&&-kT h_2(2p(1-p)(1-\delt)),
\end{eqnarray}
where as before $h_2(\cdot)$ denotes the binary entropy. By choosing $M$ large enough, we can ensure that the extracted work is arbitrarily peaked around the average given by the above equation ($M$ can be bounded using the results of \cite{aberg2013truly}). This ensures that when we need to repump the reference, we actually have enough work to invest to do it. The repumping costs $\bar{M}+s\sigma_M^{4/3}$ units of extracted work and the cost $W_E$ of the measurement is independent from $M$. Hence, the net gain per processed copy of $\ket{\psi}$ is given by
\begin{equation}\small
\langle \tilde{W} \rangle - \frac{(\bar{M}+s\sigma_M^{4/3})+W_E}{M}=\langle W \rangle -O\left(M^{-1/3}\right),
\end{equation}\normalsize
where
\begin{eqnarray}
\nonumber
\langle W \rangle &=& \Delta F(\ket{\psi}) - 2p(1-p)(1-\delt) \\
&&-kT h_2(2p(1-p)(1-\delt)).
\end{eqnarray}
Therefore, the deficit per copy scales as $M^{-1/3}$ and by choosing $M$ large enough it can be made arbitrarily small. Moreover, the previous equation gives us the relation between the quality of the reference and the average extracted work, showing that $\langle {W} \rangle \rightarrow \Delta  F(\ket{\psi})$ as $\delt \rightarrow 1, M\rightarrow \infty$, $M(1-\delt) \rightarrow 0$.

We conclude that it is possible, with arbitrarily large success probability, to extract an amount of work arbitrarily close to the free energy change from a pure state with coherence in energy eigenbasis, while processing it individually and properly taking account of all the resources used, i.e., ensuring arbitrarily exact repeatability.

\section{Extracting work with perfect repeatability and bounded thermal machines}
\label{sec:perfect}

In the previous section we have shown how to extract as work all free energy of a pure quantum state with coherence. However we allowed for
\begin{enumerate}
\item The limit case of an unbounded thermal machine, $\<\bar{\Delta}\> \rightarrow 1$.
\item An asymptotic protocol individually processing a large number $M$ of copies of the system.
\end{enumerate}
These assumptions may be too strong if the reference itself is a microscopic system involved in the thermodynamic processing and exclude the applicability to single-shot scenarios. What if we only want to process a small number of systems? This requires us to go beyond the results of the previous section.

Moreover, even if we only want to release the first of the two assumptions, i.e., put a bound on the coherence properties of the reference, we are still left with open questions. In this case the general result stated by Theorem \ref{theorem2} is applicable, however the work extraction protocols presented always entail a failure probability $1- p_{\rm{succ}}$ that can lead to a complete destruction of the coherent properties of the reference. Even if this probability is relatively small, we may not be willing to take this risk. Also, the reference inevitably deteriorates, even if by a small amount bounded by Eq.~\eqref{eq:fundamental}. A crucial question is then: are there work extraction protocols with $\langle \bar{\Delta} \rangle <1$ such that $\delta \langle \bar{\Delta} \rangle = 0$ and $p_{\rm{succ}}=1$? In other words, can we extract work from coherence using a protocol that \emph{never} fails and gives back the thermal machine with \emph{exactly} the same quality parameters, even if the reference is bounded?

In this section we construct such protocols for both average and single-shot work extraction. These ensure perfect repeatability, but the price we pay is that the average amount of extracted work is strictly smaller than the free energy difference and it is only possible for $\<\bar{\Delta}\>$ above a certain threshold value $\bar{\Delta}_{\rm{crit}}$. In the case of single-shot work extraction we show similarly that there exists a threshold over which the reference allows us to outperform the single-shot protocol with no coherence. For clarity of the discussion, we focus on the paradigmatic case of the class of states $\ket{\gamma}$ introduced in Eq.~\eqref{eq:ketgamma}. 

\subsection{Average work extraction}

In absence of an external source of coherence no work can be extracted from the state $\ket{\gamma}$ on average \cite{skrzypczyk2013extracting,lostaglio2015description}. However, if we allow for a repeatable use of the reference, positive work yield can be obtained. In order to achieve this, during step (\ref{step2}) of the protocol we perform average work extraction from the state $\D(\rho_S')$ specified by Eq.~\eqref{eq:dephased}. As $\mathcal{D}(\rho'_S)$ is diagonal in the energy eigenbasis, the results of \cite{aberg2013truly,skrzypczyk2013extracting} apply. Therefore, the average work yield is given by the free energy difference $\Delta F(\mathcal{D}(\rho'_S))$. To ensure perfect repeatability we repump the reference at each run, so that if $\bra{0}\rho_R\ket{0} =0$, then $\bra{0}\rho''_R\ket{0}=0$, and the reference quality parameters do not change. The repumping requires a unit of work, so that the work extracted on average during one run of the protocol is
\begin{equation}
\label{eq:averageworkyield}
\langle W \rangle = q+ kT (\log Z_S - h_2(q)) -1.
\end{equation}
The connection between the properties of the reference and the work yield is given by Eq.~\eqref{eq:averageworkyield} together with Eq.~\eqref{eq:q} (where \mbox{$R_{00}=0$} and $p=r$).

\begin{figure}[t!]
\centering
\includegraphics[width=0.9\columnwidth]{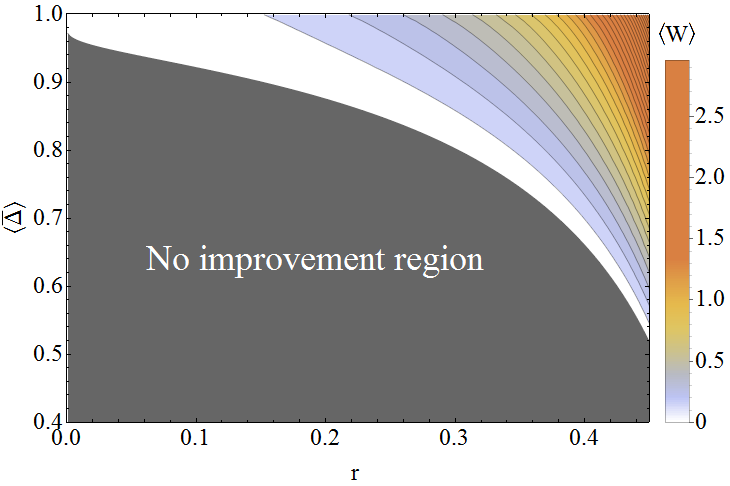}
\caption{\label{fig:average} \textbf{Coherence boost to average work extraction from $\ket{\gamma}$}. Work can be unlocked from the coherence of the system using a thermal machine that never deteriorates. The quality of the machine, measured by $\langle \bar{\Delta} \rangle$, must be bigger than some threshold value (boundary of the grey region) to ensure $\langle W\rangle >0$ ($r$ denotes the thermal occupation of the excited state, $r=(1+e^{1/kT})^{-1}$). Over the threshold, the higher the quality the greater is the average work yield from quantum coherence.}
\end{figure}

In Fig.~\ref{fig:average} we show how much work $\langle W \rangle $ can be unlocked through our protocol as a function of the quality of the reference $\langle \bar{\Delta} \rangle$ and the thermal occupation $r$ of the excited state. The graph shows that the quality of the reference needs to be above a certain threshold in order to get positive average work yield. As expected, the advantage is the most significant for high $r$, because the states $\ket{\gamma}$ and $\gamma_S$ differ most in this case or, in other words, the amount of coherence to be unlocked is higher.

As already mentioned in Sec.~\ref{sec:two}, in the asymptotic regime of individually processing large number of copies of $\ket{\gamma}$, the fluctuations in the work yield,  Eq.~\eqref{eq:averageworkyield}, become negligible. Notice, however, that even if $\langle W \rangle >0$ we may not be able to perform step~(\ref{step3}) every time, as the fluctuations around the average mean that we will not always have enough work to invest in the repumping. To resolve this problem we can follow a strategy analogous to the case of unbounded reference. That is, we repump after having extracted work $M$ times, where $M$ is sufficiently large to neglect the fluctuations around $\langle W\rangle $.\footnote{One can think of alternative protocols as well, in which at every repetition we toss a coin to decide if we repump the reference or not. We do not delve into this, but we expect to find similar results.} The protocol will be repeatable up to an arbitrarily small probability of failure, if the support of the reference initially starts high enough in the energy ladder. It is important to stress, however, that the ``failure'' in this case does not entail a destruction of the coherence properties of the reference, as in Sec.~\ref{sec:classical}. It only requires the investment of extra work in order to ensure perfect repeatability.

\subsection{Single-shot work extraction}

\begin{figure}[t!]
\centering
\includegraphics[width=0.9\columnwidth]{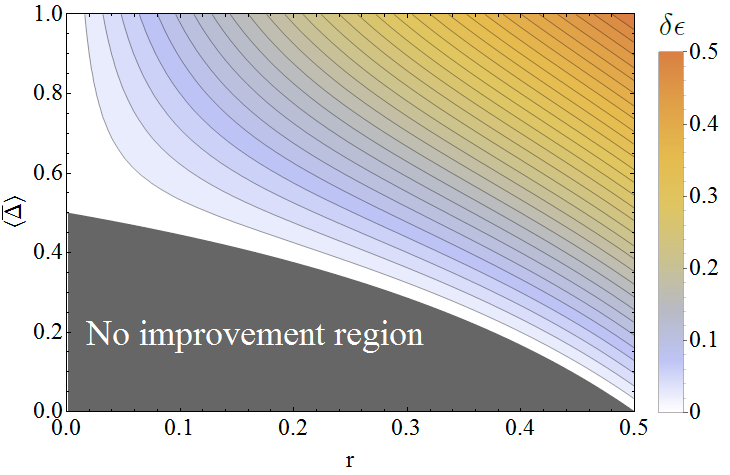}
\caption{\label{fig:singleshot} \textbf{Coherence boost to single-shot work extraction from $\ket{\gamma}$}. A thermal machine, used in a repeatable way, can exploit the quantum coherence of the system to decrease the failure probability in single-shot work extraction. Similarly to the case of average work extraction, the quality of the machine $\<\bar{\Delta}\>$ must be over some threshold value to lead to any improvement ($r$ denotes the thermal occupation of an excited state). As $\langle \bar{\Delta} \rangle$ increases, the failure probability decreases from $r$ to $r - \delta \epsilon$, down to zero, i.e., to the point when the single-shot work extraction from the pure quantum state $\ket{\gamma}$ becomes fully deterministic.}
\end{figure}

Finally, we proceed to a fully quantum and single-shot protocol for an individual quantum state. This version of the protocol does not assume possessing an unbounded reference nor it requires asymptotic number of runs. In absence of an external source of coherence we can perform \mbox{$\epsilon$-deterministic} work extraction from $\ket{\gamma}$ -- as $\ket{\gamma}$ is indistinguishable from $\gamma_S$, the results of \cite{aberg2013truly} apply. This means that we can extract $kT \log Z_S$ work with failure probability $r$ or $1+kT \log Z_S$ with failure probability $1-r$. We now show that exploiting the reference in a perfectly repeatable way the failure probability for extracting $k T \log Z_S$ can be decreased -- and the higher the quality of the reference, the stronger the improvement.

During step (\ref{step2}) of the protocol we perform \mbox{$\epsilon$-deterministic} work extraction from the state $\D(\rho_S')$ specified by Eq.~\eqref{eq:dephased}, in accordance with Assumption~\ref{assumption3}. With probability $q$ we extract \mbox{$1+kT\log Z_S$} work and with probability $1-q$ our protocol fails. As we need one unit of work to repump the reference [see Eq.~\eqref{eq:repumping}], the net gain is $k T \log Z_S$. When the protocol fails (with probability $q$), the reference is returned in the state $\rho_R'$ and one has to invest one unit of work to ensure repeatability. 

In Fig.~\ref{fig:singleshot} we present the decrease $\delta \epsilon$ in the failure probability $\epsilon$ achieved by our protocol as compared to work extraction from $\gamma_S$. We see that if the quality of the reference is high enough, the coherence content of $\ket{\gamma}$ can be exploited to provide an advantage in the work extraction. In the limit of a very high quality (unbounded) reference, $\langle \bar{\Delta} \rangle \rightarrow 1$, the failure probability can be sent to zero, i.e., the work extraction from $\ket{\gamma}$ becomes deterministic.

\section{Conclusions}

In this work we have addressed the following question: how much work can be extracted from a state that is a superposition of energy eigenstates? We argued that this question, within the currently developed theory of thermodynamics of individual quantum systems, is a subtle issue. We showed that the optimal coherence to work conversion can be obtained only in the limit of accessing a reference system with unbounded coherence resources. Although no real reference is unbounded (in the same way in which no heat engines is ideal), we can get arbitrarily close to the limit by means of a sequence of bounded thermal machines.

The access to arbitrarily large resources should be questioned in the regime under study. Generally speaking, recovering traditional thermodynamical results requires extra assumptions, which all entail some notion of ``classicality'', so to effectively make coherence negligible: neglecting the energy fluctuations due to superposition; assuming the existence of a source of coherence that experiences essentially no back-reaction; collectively operating on infinitely many copies of the system. 

When the ``classicality'' assumptions are dropped one after the other, the results are quantitatively different from the thermodynamics of incoherent systems. Nevertheless, we find that the coherence between energy levels can still enhance the performance of work extraction protocols. There exist perfectly repeatable processes extracting on average a larger amount of work that could be extracted in the absence of coherence; and single-shot protocols in which coherence improves the success probability of work extraction. Although these protocols are better than the correspondent incoherent ones, they do not achieve the performance reached in the classical limit.

We also point out that while dealing with microscopic systems, the accounting of all the resources involved in thermodynamic processes becomes a crucial and non-trivial task. In this regard, we underline the importance of accounting for the resources which make up a thermal machine, and the concept of repeatability, that essentially captures the idea of using these extra resources without degrading them. In particular, the considerations here
suggest that a full theory of thermodynamics in the quantum regime will require a better understanding of the accounting of coherence resources, including those
found in the thermal machine. Some laws which place restrictions on coherence have been introduced in \cite{brandao2013second,lostaglio2015description,cwiklinski2014limitations,narasimhachar2014low,lostaglio2015quantum}, but we are still far from having a full understanding.
We hope to have convinced the reader that the question of the role of quantum coherence in thermodynamic considerations does not admit an easy and immediate answer, and that it is only by appropriately incorporating it into the theoretical framework that we can explore truly quantum mechanical effects.

\bigskip

\textbf{Acknowledgements:} We thank Terry Rudolph, Tony Short and Philipp Kammerlander for helpful discussions. We would also like to thank Janet Anders for her useful comments on a draft of this paper. Finally, we are grateful for constructive criticism and suggested changes provided by an anonymous referee. KK and ML are supported by EPSRC and in part by COST Action MP1209. DJ is supported by the Royal Society. JO is supported by an EPSRC Established Career Fellowship.

\section*{A\lowercase{ppendix} A - C\lowercase{ollective processing regime}}

In the collective processing regime work can be effectively unlocked from coherence. This is achieved by processing many copies of a system state $\rho_S$ collectively and extracting work from relational degrees of freedom that live in decoherence-free subspaces \cite{bartlett2007reference,brandao2011resource,skrzypczyk2013extracting, lostaglio2015description}. The intuitive explanation is that one copy of a state $\rho_S$ with coherence can act as a reference for the other one, and we have \mbox{$\D(\rho_S^{\otimes 2})\neq \D(\rho_S)^{\otimes 2}$}. In the case of finite number of copies $\rho_S^{\otimes N}$ a non-zero amount of work is unlocked from the coherences, and in the limit of processing collectively infinitely many independent and identically distributed (i.i.d.) copies, the amount of work per copy that can be extracted deterministically equals $F(\rho_S) - F(\gamma_S)$. 

Instead of collectively processing many copies of a system, one may also consider a black-box device $B$ that takes in individual quantum systems $\rho_S$ and at \emph{each round} returns a thermalized state $\gamma_S$ and an average amount of work equal to $F(\rho_S) - F(\gamma_S)$. From outside the box it seems we are dealing with a work extraction protocol that individually processes each state. However, the devious way in which the box achieves this is the following: 
\begin{enumerate}
\item The box $B$ contains a large quantum memory consisting of $N\gg 1$ copies of incoherent quantum states $\sigma_B^{\otimes N}$, for which $F(\sigma_B) = F(\rho_S)$. 
\item Every time the box takes in a single copy $\rho_S$ it swaps this state into memory and instead performs work extraction on one copy of $\sigma_B$. Hence it outputs on average $F(\rho_S) - F(\gamma_S)$ and a thermalised state $\gamma_S$.
\item After $N$ uses its memory is filled with the coherent states $\rho^{\otimes N}_B$ and so it does large-$N$ collective processing and restores to $\sigma_B^{\otimes N}$ with costs growing only sublinearly with $N$ \cite{brandao2011resource}.
\end{enumerate}
Although from the outside of the box this is identical to the individual processing regime, the collective, relational processing of coherence is ``hidden'' in the quantum memory.

\section*{A\lowercase{ppendix} B - A\lowercase{verage energy conservation does not explicitly model energy fluctuations}}

Consider a system and an ancilla described by Hamiltonians $H_S$ and $H_A$, and prepared in states $\rho_S$ and $\rho_A$, respectively. Assume also that the initial state of the system $\rho_S$ has coherence between energy eigenspaces. Now consider a joint energy-conserving unitary $U$, i.e., $[U,H_S+H_A]=0$, inducing the following evolution:
\begin{equation*}
U(\rho_S\otimes\rho_A)U^{\dagger}=\rho_S'\otimes\rho_A',
\end{equation*}
so that the final state of the system $\rho_S'$ has no coherence in the energy eigenbasis and $\rho'_S = V \rho_S V^{\dag}$ for some unitary $V$ that conserves average energy. The uncertainty of an energy measurement on $\rho'_{A}$ can be decomposed as \cite{korzekwa2014quantum}:
\begin{equation}
\label{eq:decomposition}
H(\rho'_{A}) = S(\rho'_A) + A(\rho'_A),
\end{equation}
where \mbox{$A(\sigma) = S(\sigma || \mathcal{D}(\sigma))$} is the relative entropy between a state and its decohered version and $H$ is the Shannon entropy of the probability distribution of an energy measurement. Because $U$ commutes with the total Hamiltonian we have
\begin{equation*}
A(\rho_S\otimes\rho_A)=A(\rho_S'\otimes\rho_A').
\end{equation*}
As the final state of the system $\rho_S'$ has no coherence we have \mbox{$A(\rho_S'\otimes\rho_A')=A(\rho_A')$}. Using $A(\rho_S \otimes \rho_A) > A(\rho_A)$, one gets that $A(\rho'_A) > A(\rho_A)$. 
From the invariance of the von Neumann entropy under unitary transformations, $S(\rho'_A) = S(\rho_A)$. So we conclude from Eq.~\eqref{eq:decomposition}
\begin{equation*}
H(\rho'_{A}) > S(\rho_A) + A(\rho_A) = H(\rho_{A}).
\end{equation*}

\section*{A\lowercase{ppendix} C - D\lowercase{etails of the repumping stage}}

Although one could question the repumping stage described by Eq.~\eqref{eq:repumping}, given that $\Delta$ is not a unitary, we note that this is actually not a problem. This is because such operation can be realized through a joint energy-conserving unitary between a weight system in a state $\ket{1}$ and the reference in a state $\rho'_R$. The unitary is given by $V(U)$ in Eq.~\eqref{abergdynamics}, where we take $U=X$, the Pauli $X$ operator. Then
\begin{equation*}
V(X) = \sigma_- \otimes \Delta + \sigma_+ \otimes \Delta^{\dag} + \ketbra{0}{0} \otimes \ketbra{0}{0},
\end{equation*}
where $\sigma_+ = \ketbra{1}{0}$ and $\sigma_- = \ketbra{0}{1}$. As the reference has no population in the ground state, the final state of the weight system is $\ket{0}$ and the final state of the reference is given by Eq.~\eqref{eq:repumping}.

\section*{A\lowercase{ppendix} D - F\lowercase{ree energy change of the reference}}

Denote by $\Delta F_{R,n}$ the change in the free energy of the reference at the $n$-th repetition of the protocol. The total free energy change of the reference after $M$ repetitions of the protocol satisfies
\begin{equation*}
\sum_{n=1}^M \Delta F_{R,n} \leq F(\rho_R) - F(\gamma_R) \quad \forall M,
\end{equation*}
where $\gamma_R$ is the thermal state of the reference. Hence, the average change in the free energy of the reference as $M\rightarrow \infty$ is
\begin{equation*}
 \overline{\Delta F_R} :=\lim_{M \rightarrow \infty}  \frac{1}{M} \sum_{n=1}^M \Delta F_{R,n} =0.
\end{equation*}

\section*{A\lowercase{ppendix} E - D\lowercase{etails of approaching free energy limit}}

We provide here the details of the repumping protocol. We start from a generic reference state $\rho_R$ such that 
\begin{equation*}
\rm{supp}(\rho_R) \cap \rm{span}\{\ket{0},...,\ket{M}\} = \emptyset,
\end{equation*}
where $\ket{i}$ are eigenstates of the reference Hamiltonian. We impose the requirements $M p \gg 1$ and $M(1-p) \gg 1$, where $p$ is fixed by Eq.~\eqref{eq:psi}. 

We now compute the probability of the occurrence of the Kraus $A_1$ on a generic reference state $\sigma$. From Eqs.~\eqref{eq:kraus1}-\eqref{eq:kraus2} and the fact that $\Delta \Delta^{\dag} = \iden$ we obtain
\begin{equation*}
p_1(\sigma) :=\trace{}{A_1 \sigma A^{\dag}_1}= 1- 2p(1-p)(1- \trace{}{\bar{\Delta} \sigma}),
\end{equation*}
where recall that $\bar{\Delta} = (\Delta + \Delta^{\dagger})/2$. Define the state of the reference after performing work extraction on $n\geq 1$ qubits through the following recurrence formula
\begin{equation}
\label{eq:reccurence}
\rho_R^{(n)} :=A_0 \rho^{(n-1)}_R A^{\dag}_0 + A_1 \rho^{(n-1)}_R A^{\dag}_1,
\end{equation}
where $\rho_R^{(0)} = \rho_R$. Because $\rho_R$ has initially no support in the first $M$ energy levels, we can extract work from $M$ qubits before there is any overlap with the ground state. In other words, $\bra{0} \rho^{(n)}_R \ket{0} = 0, \quad \forall n \in \{1,..,M\}$.
The previous formula, together with Eq.~\eqref{eq:deltadelta}, implies that $\bar{\Delta}$ is conserved throughout the protocol. Hence we deduce that
\begin{equation*}
\trace{}{A^M_1 \rho_R A^{\dag M}_1} = p_1^M(\rho_R)=(1-2p(1-p)(1-\delt))^M.
\end{equation*} 
For notational convenience, we will now drop the explicit dependence of $p_1$ on $\rho_R$ (initial state of the reference).

Using Eq. \eqref{eq:reccurence} we have
\begin{equation}
\label{eq:rhorm}
\rho_R^{(M)} = p_1^M  \rho_{R,1} + (1-p_1^M) \mathcal{E}^{(M)}_{\rm{else}}(\rho_R),
\end{equation}
where $\mathcal{E}^{(M)}_{\rm{else}}$ contains all strings of $A_0$'s and $A_1$'s different from the string consisting only of $A_1$'s and
\begin{equation*}
\rho_{R,1} = A^M_1 \rho_R A^{\dag M}_1/p_1^M.
\end{equation*}
We can now compute $A_1^M$:
\begin{equation*}
A_1^M = \sum_{k=0}^M \binom{M}{k} p^{M-k} (1-p)^k \Delta^{\dag k}.
\end{equation*}
We see that $A_1^M$ is binomially distributed in the number of lowering operations $\Delta^{\dag}$. The average number of lowerings is $\bar{M} = M(1-p)$ and the standard deviation is $\sigma_M = \sqrt{Mp(1-p)}$. We can perform a number of repumpings as in Eq.~\eqref{eq:repumping} as detailed in Appendix C. Let us denote this operation by $\mathcal{P}$. We have chosen $M$ sufficiently large so that the confidence levels associated to $\sigma_M$ are approximately gaussian. Hence, we can repump the reference $\bar{M}+s\sigma_M^{4/3}$ times, which guarantees that the reference has arbitrarily small population in states $\ket{0}\dots\ket{M}$ with a confidence level controlled by $s>0$ and increasing with $M$. More precisely, if $P_M$ is the projector on the subspace spanned by $\{\ket{0},...,\ket{M}\}$ and $P^\perp_M = \iden - P_M$,
\begin{equation}
\label{eq:projm}
\trace{}{P^\perp_M \mathcal{P}(\rho_{R,1})} \geq {\rm erf}(s M^{1/6}/\sqrt{2}):=\mathbb{E}_s(M^{1/6}),
\end{equation}
where \mbox{$\rm{erf}(x)=\frac{2}{\sqrt{\pi}}\int_0^x\exp(-t^2)dt$} denotes the error function. Now, using Eq.~\eqref{eq:rhorm} and Eq.~\eqref{eq:projm}
\begin{equation}
\label{eq:success}
p_{\rm{succ}}:=\trace{}{P^\perp_M \mathcal{P}(\rho^{(M)}_{R})} \geq p_1^M \mathbb{E}_s(M^{1/6}).
\end{equation}
This implies that in performing the two-outcome measurement $\{P_M, P^\perp_M\}$ we would find the outcome $P^\perp_M$ with probability given by Eq.~\eqref{eq:success}. 

Performing such a measurement guarantees that the final state of the reference will have no support on a subspace spanned by $\{\ket{0}\dots\ket{M}\}$, similarly to the initial state. However, performing a selective measurement has a thermodynamic cost that we have to take into account. More precisely, such a measurement can be performed using an ancillary memory qubit system $A$ described by trivial Hamiltonian $H_A=0$. Then, taking the initial state of $A$ to be a pure state $\ket{0}$, we can perform operation on the joint reference-ancillary state described by the Kraus operators \mbox{$M_1=P^\perp_M\otimes\iden$} and $M_2=P_M\otimes\sigma_x$. This operation is energy conserving, as the Kraus operators commute with the total Hamiltonian $H_R+H_A$. Hence, it is free of thermodynamic cost. Now the projective measurement on states $\ket{0}$ and $\ket{1}$ can be performed on the ancillary memory system. Observing the result 0 will project the reference on a subspace $P_M^\perp$, whereas observing the result 1 will project the reference on $P_M$. The thermodynamic cost associated with this projective measurement is the cost of erasing the memory system afterwards. This is given by 
\begin{equation*}
W_E= kT h_2(p_{\rm{succ}}),
\end{equation*}
which can be made arbitrarily small as $p_{\rm{succ}}\rightarrow 1$. Notice that we only needed to use a classical memory to record the measurement outcome, which is not in contrast with assumption 3 of Section~\ref{sec:protocol2}. Also note that this cost has to be paid only after extracting work from $M$ copies, hence the cost per copy scales as $M^{-1}$. 

Define
\begin{equation*}
\rho''_R := \frac{P^{\perp}_M  \mathcal{P} (\rho^{(M)}_{R}) P^{\perp}_M}{\norm{P^{\perp}_M  \mathcal{P} (\rho^{(M)}_{R}) P^{\perp}_M}}.
\end{equation*}
Now, using the gentle measurement lemma \cite{winter1999coding,ogawa2002new}, Eq.~\eqref{eq:success} also implies
\begin{equation}
\label{eq:gentle}
\norm{\rho''_R - \mathcal{P}(\rho^{(M)}_{R})}\leq 2\sqrt{1-p_{\rm{succ}}}.
\end{equation}
From Eq.~\eqref{eq:gentle}, and the following characterization of the trace norm (see \cite{nielsen2010quantum})
\begin{equation*}
\norm{\rho - \sigma} = \max_{0 \leq A \leq \iden}\trace{}{A(\rho - \sigma)} 
\end{equation*}
we find that
\begin{eqnarray*}
\trace{}{\bar{\Delta} \rho''_R}  & \geq & \trace{}{\bar{\Delta} \mathcal{P} (\rho^{(M)}_{R})} - 2\sqrt{1-p_{\rm{succ}}} \\ & = &\trace{}{\bar{\Delta} \rho_{R}} - 2\sqrt{1-p_{\rm{succ}}},
\end{eqnarray*}
where the last equality comes from the fact that $\langle \bar{\Delta} \rangle$ is conserved in the protocol, up to the measurement.
The last equation can be rewritten as
\begin{equation*}
\trace{}{\bar{\Delta} (\rho_{R}-\rho''_R)}\leq 2\sqrt{1-p_{\rm{succ}}}.
\end{equation*}
Exchanging the roles of $\rho_R$ and $\rho''_R$ and introducing 
\begin{equation*}
\delta \delt = \trace{}{\bar{\Delta} \rho''_R}-\trace{}{\bar{\Delta} \rho_{R}},
\end{equation*}
we conclude
\begin{equation*}
|\delta \delt|\leq 2\sqrt{1-p_{\rm{succ}}}.
\end{equation*}
Using Eq.~\eqref{eq:success}, this bounds the maximum allowed change of the quality parameter of the reference.

\bibliography{Bibliography_thermodynamics}

\end{document}